\documentclass[12pt]{article}
\usepackage{amsmath,amsfonts,amssymb,setspace,tensor}
\usepackage[utf8]{inputenc}
\usepackage{slashed}
\usepackage[nosort]{cite}
\usepackage[table]{xcolor}
\usepackage{dsfont}
\usepackage{mathrsfs}
\usepackage{upgreek}
\usepackage{graphicx}	
\usepackage{float}
\usepackage{longtable}	
\usepackage{lipsum}
\usepackage{verbatim}
\usepackage{leftidx}	

\usepackage{bm}			
\usepackage{marginnote}	

\numberwithin{equation}{section} 

\usepackage[top=2.8cm,
			bottom=2.8cm,
			left=2.0cm,
			right=2.0cm]{geometry}
\usepackage[colorlinks=true,
			linkcolor=black,
			citecolor=black,
			urlcolor=blue,
			filecolor=black]{hyperref}

\allowdisplaybreaks

\begin{document}

\vspace{30pt}

\begin{center}


{\Large\sc Non-Null Torus Knotted Gravitational Waves 
\\
from Gravitoelectromagnetism 
\\[12pt]}

\vspace{-5pt}
\par\noindent\rule{420pt}{0.5pt}


\vskip 1cm

{\sc R. S. Facundo and I. V. Vancea}


\vspace{10pt}
{\it 
Group of Theoretical Physics and Mathematical Physics,\\
Department of Physics, Federal Rural University of Rio de Janeiro,\\
Cx. Postal 23851, BR 465 Km 7, 23890-000 Serop\'{e}dica - RJ,
Brazil}

\vspace{4pt}


{\tt\small
\href{mailto:rian.s.f@hotmail.com}{rian.s.f@hotmail.br}
}
{\tt\small 
\href{mailto:ionvancea@ufrrj.br}{ionvancea@ufrrj.br}
}


\vspace{30pt} {\sc\large Abstract} \end{center}

In this paper, we construct a non-null torus-knotted gravitational monochromatic wave solution of the linearized Einstein equations in vacuum, employing the gravitoelectromagnetic (GEM) framework by analogy with classical electrodynamics. We derive the geometric objects, including the line element, the Riemann tensor, the Ricci tensor, the Ricci scalar, and the geodesic equation for this background. Also, we investigate two properties inherent to this solution due to its GEM origin: the dual GEM potential and GEM helicity.

\noindent



\newpage

\section{Introduction}

The gravitoelectromagnetism (GEM) approach to weak gravitational fields is based on the observation that the mixed components $h_{0\nu}$ of small metric perturbations $\vert h_{\mu\nu} \vert \ll 1$ play an analogous role in the gauge-fixed linearized Einstein equations to the vector potential $A_{\mu}$ in classical electromagnetism \cite{Mashhoon:2000jq,Clark:2000ff,Kopeikin:2001dz}. The analogy between the GEM equations and Maxwell's equations is significant as it may reveal new gravitational effects and potential connections between gravitational and electromagnetic systems, such as condensed matter. For recent reviews on these topics, see \cite{Ciufolini:2010,Ruggiero:2002hz,Ummarino:2021tpz}.

A key problem where the analogy between electrodynamics and weak gravity within the GEM framework proves useful is constructing gravitational wave solutions to the linearized Einstein equations with non-trivial topology. Electromagnetic fields with Hopf and knot topologies are well-known solutions of the Maxwell equations in vacuum \cite{Trautman:1977im,Ranada:1989wc,Arrayas:2011ia}, and their properties have been examined in various contexts \cite{Arrayas:2011ci,Arrayas:2016bqn,Arrayas:2017xvo,Ranada:2017ore,Alves:2017ggb,Arrayas:2018a,Crisan:2020krc,Crisan:2020isv}. Guided by the GEM analogy, weak gravitational fields with Hopf and knot structures have been found and classified in the Weyl tensor formulation of GEM \cite{Maartens:1997fg,Thompson:2014pta,Thompson:2014owa,Sabharwal:2019ngs}. Also, there are methods to construct knotted solutions in electromagnetism using de Sitter space \cite{Lechtenfeld:2017tif,Kumar:2020xjr}.

In the standard GEM approach to gravitational systems in weak fields, where the analogy with Maxwell equations in the vector formalism is more direct, results concerning topologically non-trivial fields address more general existence problems \cite{Kopinski:2017nvp,Vancea:2017tmx,Smolka:2018rup,Alves:2018wku,Silva:2018ule,Crisan:2021pvz,Bini:2021gdb,Giardino:2021gwq} (see recent reviews \cite{Arrayas:2017sfq,Vancea:2019zdl}). Concrete examples of Hopf and knot weak gravitational fields in the standard GEM framework are scarce in the literature \cite{Kopinski:2017nvp,Alves:2018wku,Silva:2018ule,Hojman:2023vng,Grzela:2024ahc}. Furthermore, a conceptual problem arises in the standard GEM formulation, as it typically applies to the $h_{0\mu}$ components of the metric perturbation $h_{\mu\nu}$, leaving the interpretation of spatial components $h_{ij}$ and their relation to knotted gravitoelectric and gravitomagnetic fields unresolved.

In this paper, we address both issues by constructing a full perturbative metric solution of the linearized Einstein field equations using an extended GEM formulation. This approach extends the traditional GEM sector from $h_{0\nu}$ to the full metric $h_{\mu\nu}$. We focus on obtaining monochromatic gravitational waves with amplitudes carrying non-null torus knot topology information of gravitoelectric and gravitomagnetic fields, termed non-null torus knotted gravitational monochromatic waves (NNTKGMW). These fields represent important new examples of toroidal knotted gravitational waves where all components of $h_{\mu\nu}$ carry topological information. The torus knot emerges as a solution in the GEM sector $h_{0\mu}$ of the metric, with its existence ensured by the GEM-electromagnetism analogy. The extended GEM formalism allows us to extend this non-null wave solution to the spatial sector of the metric $h_{ij}$, which obey the remainder of the linearized gravity equations of motion with no electromagnetic counterparts.

For NNTKGMW solutions to exist, specific conditions must be met. First, like in electrodynamics, the fields must be in vacuum. Second, to define extended GEM fields, gauge freedom must be fixed by choosing global Cartesian coordinates on the flat background space-time. Third, to specify the torus, metric perturbations need further fixing. Generally, obtaining concrete GEM solutions for the full perturbative metric, analogous to classical electrodynamics, is challenging, despite the GEM-Maxwell equations analogy. This challenge stems from only partial metric perturbation components acting as electromagnetic potentials, with GEM's gauge conditions being more extensive than those in electrodynamics. Implementing gauge conditions ensures the existence of non-null torus knotted GEM solutions formally similar to the ones obtained for Maxwell's equations \cite{Clark:2000ff,FilipeCosta:2006fz,Costa:2009nn,Costa:2012cw}. The use of the extended GEM framework is justified by non-null torus knotted fields' time-dependent nature and the need to extend them to the spatial sector of the perturbed metric. These toroidal knot aspects are challenging in standard GEM, where $h_{ij}$ are typically zero or neglected based on metric component sizes, and gauge conditions may eliminate time-dependent potentials. The extended GEM formalism circumvents both potential complications \cite{Bakopoulos:2014exa,Bakopoulos:2016rkl}.

Importantly, NNTKGMW fields are constructed in a vacuum. Generally, GEM fields in vacuum lack direct physical meaning. Nonetheless, GEM potentials, associated with small perturbations of the flat space-time metric, carry information about the vacuum's non-trivial topology that might be interesting for quantum gravity \cite{Cho:2013kra}.

This paper is organized as follows. Section 2 briefly reviews the extended GEM approach to the linearized Einstein equations to make the paper self-contained, with particular attention to gauge potentials and gauge fixing necessary for deriving equations of motion for GEM potentials and spatial metric components. In Section 3, we present NNTKGMW solutions of the GEM field equations, derived analogously to Maxwell's equations' corresponding solutions. NNTKGMW fields are Fourier components of non-null torus knotted GEM potentials. The NNTKGMW solution is extended from GEM to the spatial metric sector using gauge equations. Section 4 explores fundamental properties of the NNTKGMW geometry by calculating relevant geometrical objects: Christoffel symbols, the Riemann tensor, and the Ricci tensor. Additionally, we derive the geodesic equation for the NNTKGMW background. Section 5 investigates two NNTKGMW background properties due to its GEM origin: the dual GEM potential, introducing the dual perturbative metric concept, and gravitomagnetic and gravitoelectric helicities. Finally, we discuss our results and outline future research problems. Throughout this work, the Minkowski metric is $\text{diag}(+, -, -, -)$, and the torus radius is taken as unity.


\section{Extended GEM Ansatz}

In this section, we review the basics of the extended GEM approach to linearized Einstein's equations. We follow \cite{Mashhoon:2003ax} for general GEM concepts and \cite{Bakopoulos:2014exa,Bakopoulos:2016rkl} for the extended GEM ansatz.

The linearized gravitational field around the Minkowski background $\eta_{\mu \nu}$ is given by the first-order expansion in the metric perturbation 
\begin{equation}
g_{\mu \nu} \left(x \right) = \eta_{\mu \nu} + h_{\mu \nu}\left(x \right)
\, ,
\label{GEM-knot-wave-lin-metric}
\end{equation}
where $\vert h_{\mu \nu} (x) \vert \ll 1$. By introducing the trace-reversed metric perturbation
\begin{equation}
\tilde{h}_{\mu \nu}=h_{\mu \nu}-\frac{1}{2} \eta_{\mu \nu} h 
\, ,
\label{GEM-knot-wave-tilde-h}
\end{equation}
where $h = \eta^{\mu \nu} h_{\mu \nu}$, the linearized Einstein equations take the form
\begin{equation}
\tilde{h}\indices{^\alpha _\mu _, _\nu _\alpha} 
+ 
\tilde{h}\indices{^\alpha _\nu _, _\mu _\alpha}
-
\eta_{\mu \nu} \tilde{h}\indices{^\alpha ^\beta _, _\alpha _\beta}
-
\partial^2 \tilde{h}_{\mu \nu}= 2 k T_{\mu \nu}
\, ,
\label{GEM-knot-wave-Einstein}
\end{equation}
where $\partial^2 = \eta^{\mu \nu} \partial_\mu \partial_\nu$, $k = 8 \pi G c^{-4}$, and $T_{\mu \nu}$ is the matter energy-momentum tensor. Equations (\ref{GEM-knot-wave-Einstein}) are invariant under the gauge transformations
\begin{align}
x^\mu & \rightarrow x^{\prime \mu}=x^\mu - \xi^\mu
\, ,
\label{GEM-knot-wave-gauge-1}
\\
h_{\mu \nu} & \rightarrow h_{\mu \nu}^{\prime} = h_{\mu \nu} + 
\xi_{\mu, \nu} + \xi_{\nu, \mu}
\, ,
\label{GEM-knot-wave-gauge-2}
\end{align}
where $\xi=\xi^{\mu} \partial_{\mu}$ is an infinitesimal vector field.

The GEM formalism is based on the observation that under certain circumstances, the $h_{0\mu}$ components of the metric perturbation tensor are analogous to the electromagnetic potential. To cast equations (\ref{GEM-knot-wave-Einstein}) into the GEM form, we pick the transverse gauge defined by
\begin{equation}
\tilde{h}^{\mu \nu}{ }_{, \nu}=0
\, ,
\label{GEM-knot-wave-gauge-transverse}
\end{equation}
in which the Einstein equations simplify to
\begin{equation}
\partial^2 \tilde{h}_{\mu \nu}=-2 k T_{\mu \nu}
\, .
\label{GEM-knot-wave-Einstein-2}
\end{equation}
As our focus is on solutions of the Einstein equations with properties similar to the electromagnetic knotted fields satisfying Maxwell's equations in vacuum, we take $T_{\mu \nu} = 0$. 

The GEM construction requires identifying the fields $\tilde{h}_{0 i}$ with the spatial components of a GEM vector potential. However, this identification leaves other components of $\tilde{h}_{\mu \nu}$ arbitrary, allowing multiple choices. Here, we adopt the extended GEM equations proposed in \cite{Bakopoulos:2016rkl}, where the GEM sector of $\tilde{h}_{\mu \nu}$ is given by
\begin{equation}
\tilde{h}_{00} (x)=\frac{\Phi (x)}{c^2}, \qquad \tilde{h}_{0 i} (x)=
\frac{A_i (x)}{c^2}
\, .
\label{GEM-knot-wave-ansatz-1}
\end{equation}
As discussed in the literature 
\cite{FilipeCosta:2006fz,Costa:2009nn,Costa:2012cw,Bakopoulos:2014exa,Bakopoulos:2016rkl,Mashhoon:2008kq}, 
the definition of the spatial metric sector $\tilde{h}_{ij}$ determines whether the GEM potential is stationary or time-dependent. The former often involves ignoring parts of the gauge equations (\ref{GEM-knot-wave-gauge-transverse}) on relative size grounds, potentially overlooking time-dependent solutions. Based on these considerations, the following extended GEM ansatz for the spatial sector of $\tilde{h}_{\mu \nu}$ was proposed in \cite{Bakopoulos:2014exa,Bakopoulos:2016rkl}
\begin{equation}
\tilde{h}_{i j} (x) =-\frac{1}{c^2} \eta_{i j} \Phi (x) + 
\frac{2}{c^4} \gamma_{i j} (x)
\, .
\label{GEM-knot-wave-ansatz-2} 
\end{equation} 
Here, $\gamma_{i j} = \gamma_{j i}$ and $\mathrm{Tr} (\gamma_{i j})=0$. The denominator $c^4$ is necessary to describe slow-moving systems, e.g., a non-relativistic fluid distribution \cite{Mashhoon:2003ax}, and to maintain the relative magnitude of terms in GEM equations. Plugging equations (\ref{GEM-knot-wave-ansatz-1}) and (\ref{GEM-knot-wave-ansatz-2}) into the transverse gauge (\ref{GEM-knot-wave-gauge-transverse}), we obtain the following gauge conditions on components
\begin{align}
\partial_0 \Phi + \partial_i A^i & = 0
\, ,
\label{GEM-knot-wave-mu0}
\\
- \partial^i \Phi + \partial_0 A^i  & =
- \frac{2}{c^2} \partial_j \gamma^{i j} 
\, .
\label{GEM-knot-wave-mui}
\end{align}
We recognize the Lorentz gauge in the first equation (\ref{GEM-knot-wave-mu0}) and the definition of the gravitoelectric field in the second equation. The equations of motion (\ref{GEM-knot-wave-Einstein-2}) for components read
\begin{equation}
\partial^2 \Phi=0 \, ,
\qquad
\partial^2 A^i=0 \, ,
\qquad
\partial^2 \gamma_{i j}=0  
\, .
\label{GEM-knot-wave-eq-comp}
\end{equation}
The set of equations (\ref{GEM-knot-wave-mu0}), (\ref{GEM-knot-wave-mui}) and (\ref{GEM-knot-wave-eq-comp}) are analogous to the Lorentz gauge and the homogeneous wave equation of the electromagnetic potential. The other two equations that $\gamma_{ij}$ must obey are typical to linearized gravity and do not have an electromagnetic correspondence in the vectorial formulation\footnote{However, see \cite{Bakopoulos:2014exa} for an attempt to provide a tensor formulation of electromagnetism similar to linearized gravity.}.

The previous considerations lead to the familiar definitions of the gravitoelectric and gravitomagnetic fields in terms of the GEM potentials in the vector formulation
\begin{equation}
\mathbf{E} = -\mathbf{\nabla} \Phi -\partial_t \mathbf{A} \, , \qquad 
\mathbf{B} = \nabla \times \mathbf{A}
\, .
\label{GEM-knot-wave-fields-def}
\end{equation}
It should be noted that in the presence of matter, an additional numerical factor of 4 should be included in the definition of the GEM potentials to reproduce Maxwell's equations. However, this factor is irrelevant in vacuum. Consequently, the GEM equations in vacuum are
\begin{align}
\nabla \cdot \mathbf{E} & =0 
\, , 
\quad 
\nabla \times \mathbf{E} = 
- \partial_t \mathbf{B}
\, ,
\label{GEM-knot-wave-GEM-1}
\\
\nabla \cdot \mathbf{B} & = 0
\, , 
\quad 
\nabla \times \mathbf{B} = c^{-2} \partial_t \mathbf{E}
\, .
\label{GEM-knot-wave-GEM-2}
\end{align} 
Equations (\ref{GEM-knot-wave-GEM-1}) and (\ref{GEM-knot-wave-GEM-2}) describe the dynamics of the GEM fields in the absence of sources, where the potentials are the components $\tilde{h}_{0\mu}$ of the metric perturbations in the gauge described above. These equations contain no information about $\gamma_{ij}$. Therefore, they must be supplemented by the gauge constraints and the equations of motion for $\gamma_{ij}$, as outlined above.

\section{NNTKGMW Solution}

GEM fields with non-null torus-knotted topology are those solutions to equations (\ref{GEM-knot-wave-GEM-1}) and (\ref{GEM-knot-wave-GEM-2}), characterized by the field lines of $\mathbf{E}$ and $\mathbf{B}$ winding around the cycles of a two-torus $\mathbb{T}^2 \simeq S^1 \times S^1$. Such fields were discovered in electromagnetism some time ago \cite{Trautman:1977im,Ranada:1989wc,Arrayas:2011ia}. Here, we construct NNTKGMW solutions of linear gravity (\ref{GEM-knot-wave-mu0}) and (\ref{GEM-knot-wave-mui}), which are the wave equations for the GEM potentials and the spatial metric components.

These torus-knotted fields are obtained by closely following the analogy with Maxwell's equations. As in electrodynamics, the topological character of NNTKGMW is encoded in the wave amplitudes rather than phases, determined by the non-null torus-knotted gravitoelectric and gravitomagnetic fields $\mathbf{E}$ and $\mathbf{B}$ that satisfy the GEM equations (\ref{GEM-knot-wave-GEM-1}) and (\ref{GEM-knot-wave-GEM-2}) given above.

To obtain non-null torus-knotted fields, the spatial orientation of $\mathbb{T}^2$ must be defined, equivalent to fixing the Cartesian coordinates $\{ x^\mu \}$ in the Minkowski space-time, with the additional condition $\Phi = 0$. This gauge conditions, which we call knot gauge, ensures the existence of non-null toroidal solutions to Maxwell's equations in electrodynamics and must be imposed in GEM as well.

Let us start with the monochromatic wave solution of the potential equations (\ref{GEM-knot-wave-eq-comp}), which can be written as
\begin{equation}
A^j (x;k )= 
a^j (\mathbf{k}) e^{-i k x} 
\, ,
\label{GEM-knot-wave-sol-1}
\end{equation}
where $kx=\eta_{\mu \nu}k^\mu x^\nu$, $\mathbf{k} = \{ k^i \}_{i=1,2,3}$, and $kk = \omega^2 - \mathbf{k}^2 = 0$ is the dispersion relation. The information about the non-null torus knot is contained in the amplitude $\mathbf{a} (\mathbf{k}) = \{ a^i (\mathbf{k}) \}_{i=1,2,3}$ at the initial time $x^0 = 0$. Since $A^j (x;k )$ are the components of the GEM potential, their amplitudes can be written in terms of the Fourier coefficients $\mathcal{E}^j (\mathbf{k} )$ and $\mathcal{B}^j (\mathbf{k} )$ of $\mathbf{E}(x^0 = 0, \mathbf{x})$ and $\mathbf{B}(x^0 = 0 , \mathbf{x})$, respectively, whose field lines wind around torus cycles at the initial time.

The relation between $a^i (\mathbf{k})$, $\mathcal{E}^j (\mathbf{k} )$ and $\mathcal{B}^j (\mathbf{k} )$ can be derived from equations (\ref{GEM-knot-wave-fields-def}) by applying the Fourier transform. After some calculations, we obtain
\begin{equation}
a^j (\mathbf{k})= \frac{i}{2 c \vert \mathbf{k} \vert }
\left( \mathcal{E}^j (\mathbf{k}) - \frac{c}{\vert \mathbf{k} \vert }
\varepsilon\indices{^j _p _q}
k^p \mathcal{B}^q (\mathbf{k})
\right)
\, .
\label{GEM-knot-wave-sol-2}
\end{equation}
To derive the components of the amplitude $\mathbf{a}(\mathbf{k})$, we exploit the analogy between equations (\ref{GEM-knot-wave-GEM-1}) and (\ref{GEM-knot-wave-GEM-2}) and the electromagnetic field. This provides the following non-null torus knot solution to the GEM equations in vacuum \cite{Arrayas:2011ia,Crisan:2021pvz}
\begin{align}
\mathbf{E} (x^{0} = 0, \mathbf{x}) & = 
\frac{4}
{
\pi 
\left( 
1 + \vert \mathbf{x} \vert^2 
\right)^3
}
\begin{bmatrix}
    l\left[ 
        (x^{1})^2 - (x^2)^2 - (x^3)^2 + 1   
    \right] 
    \\
    2\left(
        l x^{1} x^{2} - s x^{3} 
    \right)
    \\
    2\left(
        l x^{1} x^{3} + s x^{2} 
    \right)
\end{bmatrix}
\, ,
\label{GEM-knot-wave-E-non-null-knot}
\\
\mathbf{B} (x^{0} = 0, \mathbf{x}) & = 
\frac{4}
{
\pi 
\left( 
1 + \vert \mathbf{x} \vert^2 
\right)^3
}
\begin{bmatrix}
    2\left( 
        m x^{2} - n x^{1} x^{3}     
    \right) 
    \\
    - 2\left(
        m x^{1} - n x^{2} x^{3} 
    \right)
    \\
    n \left[
        (x^{1})^2 + (x^{2})^2 - (x^{3})^2 + 1
    \right]
\end{bmatrix}
\, .
\label{GEM-knot-wave-B-non-null-knot}
\end{align}
Here, $(m,n)$ and $(l,s)$ are pairs of coprime integers, which can be, in principle, arbitrary. Therefore, the fields $\mathbf{E}$ and $\mathbf{B}$ are generally non-null. The pair $(m,n)$ describes a gravitomagnetic field whose field lines wind around the torus $n$ times around one cycle and $m$ times around the other cycle at the initial time $x^{0} = 0$. A similar interpretation holds for the pair $(l,s)$, describing a gravitoelectric field whose field lines wind around the two cycles of $\mathbb{T}^2$. Standard calculations yield the following Fourier coefficients of the fields given in equations (\ref{GEM-knot-wave-E-non-null-knot}) and (\ref{GEM-knot-wave-B-non-null-knot}) above
\begin{align}
\boldsymbol{\mathcal{E}}(\mathbf{k}) & = 
\frac{e^{-\vert \mathbf{k} \vert }}{\sqrt{2 \pi} \vert \mathbf{k} \vert}
\begin{bmatrix}
    l
    \left[ (k^2)^2+(k^3)^2 \right]
    \\
    - l k^1 k^2 - i s \vert \mathbf{k} \vert k^3
    \\
    - l k^1 k^3 + i s \vert \mathbf{k} \vert k^2
\end{bmatrix}
\, ,
\label{GEM-knot-wave-E-Fourier-non-null-knot}
\\
\boldsymbol{\mathcal{B}}(\mathbf{k}) & = 
\frac{e^{-\vert \mathbf{k} \vert }}{\sqrt{2 \pi} \vert \mathbf{k} \vert}
\begin{bmatrix}
    n k^1 k^3  + i m \vert \mathbf{k} \vert k^2
    \\
    n k^2 k^3 - i m \vert \mathbf{k} \vert k^1
    \\
    - n
    \left[ (k^1)^2+(k^2)^2 \right]
\end{bmatrix}
\, .
\label{GEM-knot-wave-B-Fourier-non-null-knot}
\end{align}
By plugging equations (\ref{GEM-knot-wave-E-Fourier-non-null-knot}) and (\ref{GEM-knot-wave-B-Fourier-non-null-knot}) into equation (\ref{GEM-knot-wave-sol-2}), we obtain the amplitude $\mathbf{a} (\mathbf{k})$ of the potential $\mathbf{A}(x)$. Substituting this further into equation (\ref{GEM-knot-wave-sol-1}) leads to the following result
\begin{equation}
\mathbf{A} (x;k ) = 
\frac{ i e^{-\vert \mathbf{k} \vert } e^{-ikx}}{2 c \sqrt{2 \pi} \vert \mathbf{k} \vert^2}
\begin{bmatrix}
    l
    \left[ (k^2)^2+(k^3)^2 \right] - c n \vert \mathbf{k} \vert k^2
    +i c m k^1 k^3
    \\
    - l k^1 k^2 + c n \vert \mathbf{k} \vert k^1
    - i \left[ s \vert \mathbf{k} \vert k^3 - c m k^2 k^3 \right]
    \\
    - l k^1 k^3 
    + i \left[ s \vert \mathbf{k} \vert k^2 - c m \left[ (k^1)^2+(k^2)^2 \right] \right]
\end{bmatrix}
\, .
\label{GEM-knot-wave-sol-3}
\end{equation}

Equation (\ref{GEM-knot-wave-sol-3}) represents the components of the GEM sector $\tilde{h}_{0i}$ of metric perturbation with $k_i A^i (x;k) = 0$. However, the GEM equations do not provide any information on the spatial sector $\tilde{h}_{ij}$. Nevertheless, in the knot gauge, $\tilde{h}_{ij} = 2 \gamma_{ij}$ are not independent of $\tilde{h}_{0i}$, as seen from the gauge equation (\ref{GEM-knot-wave-mui}). These observations allow us to determine the components $\gamma_{ij}$ of the metric tensor. Since we are discussing a single frequency wave and $\gamma_{ij}$ satisfies the wave equation (\ref{GEM-knot-wave-eq-comp}), we make the following ansatz
\begin{equation}
\gamma_{ij} (x;k )= 
\upgamma_{ij} (\mathbf{k}) e^{-i k x} 
\, .
\label{GEM-knot-wave-sol-gamma}
\end{equation}
From equation (\ref{GEM-knot-wave-eq-comp}) and by imposing the symmetry condition $\gamma_{ij} = \gamma_{ji}$, we obtain
\begin{equation}
\gamma_{ij} (x;k )= 
k_{(i} a_{j)}(\mathbf{k})  \, \frac{e^{-i k x}}{k_0}  
\, .
\label{GEM-knot-wave-sol-gamma-k}
\end{equation}
Here, the brackets denote symmetrization $a_{(i}b_{j)} = (a_i b_j + a_j b_i)/2$. Note that the right-hand side of equation (\ref{GEM-knot-wave-sol-gamma-k}) was obtained using the dispersion relation for plane waves. The tracelessness of $\gamma_{ij}$ is guaranteed by the transversality of the gravitoelectromagnetic wave.

This concludes the construction of the NNTKGMW fields. To summarize, equations (\ref{GEM-knot-wave-sol-3}) and (\ref{GEM-knot-wave-sol-gamma-k}) represent a family of monochromatic gravitational waves parameterized by the integers $(m,n,l,s)$, whose components $\tilde{h}_{0\mu}$ are the GEM potentials of the GEM torus knot fields of the particular form specified by the ansatz (\ref{GEM-knot-wave-ansatz-1}) and the three-dimensional components given by equation (\ref{GEM-knot-wave-ansatz-2}) above.

\section{NNTKGMW Geometry}

In the last section, we established a flat space-time in which there are gravitational waves $\tilde{h}_{\mu \nu} = \{ \tilde{h}_{00}, \tilde{h}_{0i} = \tilde{h}_{i0}, \tilde{h}_{ij} \} \sim \{ 0, 2 A_i , 2 \gamma_{ij} \}$ whose amplitudes contain information about the non-null torus knot of typical radius one at initial time $x^0 = 0$. In principle, this information could be inherited by geometrical objects constructed from the metric. To see how this comes about, let us explicitly calculate these objects in NNTKGMW geometry. 

\subsection{Real Line Element}

The NNTKGMW fields obtained in the previous section are described by complex functions. To restrict the discussion to real space-time backgrounds, we consider only the real part of the GEM potentials $A_i$ and $\gamma_{ij}$ given by
\begin{align} 
A^{i}_R (x;k ) 
& =
\frac{e^{- \vert \mathbf{k} \vert}}{2 \sqrt{2 \pi} \vert \mathbf{k} \vert^2}
\left[
f^{i}(\mathbf{k}) \sin (kx) + g^{i}(\mathbf{k}) \cos (kx)
\right]
\, ,
\label{GEM-knot-wave-sol-A-real}
\\
\left[ \mathrm{\gamma}_R \right]_{ij} (x;k )
& = 
\frac{e^{- \vert \mathbf{k} \vert}}{2 k_0 \sqrt{2 \pi} \vert \mathbf{k} \vert^2}
\left[
k_{(i} f_{j)} (\mathbf{k}) \sin (kx)
+
k_{(i} g_{j)} (\mathbf{k}) \cos (kx)
\right]
\, .
\label{GEM-knot-wave-sol-gamma-real}
\end{align}
Here, the constant three-dimensional vectors $\mathbf{f}(\mathbf{k})$ and $\mathbf{g}(\mathbf{k})$ depend only on the wave vector $\mathbf{k}$ and the winding numbers $(m,n,l,s)$, and are given by the following relations
\begin{align}
\mathbf{f} (\mathbf{k} ) 
& = \frac{1}{c}
\begin{bmatrix}
    l
    \left[ (k^2)^2+(k^3)^2 \right] - c n \vert \mathbf{k} \vert k^2
    \\
    - l k^1 k^2 + c n \vert \mathbf{k} \vert k^1
    \\
    - l k^1 k^3 
\end{bmatrix}
\, ,
\label{GEM-knot-vector-f}
\\
\mathbf{g} (\mathbf{k} ) 
& = \frac{1}{c}
\begin{bmatrix}
    - c m k^1 k^3  
    \\
    s \vert \mathbf{k} \vert k^3 - c m k^2 k^3
    \\
    - s \vert \mathbf{k} \vert k^2 + c m \left[ (k^1)^2+(k^2)^2 \right]
\end{bmatrix}
\, .
\label{GEM-knot-vector-g}
\end{align}
By inverting the trace-reversed metric relation (\ref{GEM-knot-wave-tilde-h}), the space-time line element corresponding to the NNTKGMW fields $A^{i}_R (x;k )$ and $\left[ \gamma_R\right] _{ij} (x;k )$ can be written as 
\begin{align}
d s^2 (k)  
& = 
(dx^0)^2
+  
\frac{e^{- \vert \mathbf{k} \vert}}{c^2 \sqrt{2 \pi} \vert \mathbf{k} \vert^2}
\left[
f_i (\mathbf{k}) \sin (kx) + g_i (\mathbf{k}) \cos (kx)
\right]
dx^0 dx^i
\nonumber
\\
& +
\left[\eta_{i j} + 
\frac{e^{- \vert \mathbf{k} \vert}}{c^4 \sqrt{2 \pi} k_0 \vert \mathbf{k} \vert^2}
\left[
k_{(i} f_{j)} (\mathbf{k}) \sin (kx)
+
k_{(i} g_{j)} (\mathbf{k}) \cos (kx)
\right]
\right] d x^i d x^j
\, .
\label{GEM-knot-wave-metric}
\end{align}
It is understood that $d s^2 (k)$ also depends on $(m,n,l,s)$, which are not explicitly written to avoid index cluttering. Equation (\ref{GEM-knot-wave-metric}) describes a family of space-time metrics parametrized by seven real parameters corresponding to the components of the wave vector $\mathbf{k}$ and the winding numbers $(m,n,l,s)$, which we call a NNTKGMW background.

The terms from the line element given in equation (\ref{GEM-knot-wave-metric}) have different magnitudes, as the coefficients of $dx^0 dx^i$ are $A_i/c^2$, while the coefficients of $dx^i dx^j$ are given by $\gamma_{ij}/c^4$. Therefore, the NNTKGMW metric exhibits a structure similar to the higher-order corrections seen in the extended GEM ansatz from equations (\ref{GEM-knot-wave-ansatz-1}) and (\ref{GEM-knot-wave-ansatz-2}) above. Specifically, the GEM component of the NNTKGMW metric represents a first-order perturbation, whereas the spatial component of NNTKGMW corresponds to a second-order perturbation. Nevertheless, as discussed in the previous section, these components should be considered simultaneously. One important consequence is that the properties of the gravitational field and the dynamics of moving bodies within it are well approximated by the weak-field and slow-motion regime.

\subsection{Curvature of NNTKGMW Background}

The geometric properties of spacetime, the effects of gravity, and the relationship between curvature and matter-energy dynamics are provided by the Riemann tensor, the Ricci tensor, and the Ricci scalar 
\cite{Misner:1973prb}. Let's proceed to the calculation of these objects for the NNTKGMW background.

\subsubsection{The Riemann Tensor}

The linearized Riemann tensor for the perturbative metric can be expressed in terms of metric components as follows
\begin{equation}
R_{\rho\sigma\mu\nu} = \frac{1}{2} \left( h_{\rho\nu, \mu \sigma} +  h_{\sigma\mu , \nu \rho} - h_{\rho\mu , \nu \sigma} - h_{\sigma\nu , \mu \rho} \right)
\, .
\label{GEM-knot-wave-Riemann-tensor}
\end{equation}
Again, the inverse of relation (\ref{GEM-knot-wave-tilde-h}) has to be used on the right-hand side of equation (\ref{GEM-knot-wave-Riemann-tensor}). After some extended but standard algebra, we obtain the non-vanishing components of the Riemann tensor
\begin{align}
R_{0 i j k} 
& = 
-\frac{2 e^{-|\mathbf{k}|} k_i}{c^2 \sqrt{2 \pi} |\mathbf{k}|^2}
\left[ 
    k_{[j} f_{k]} (\mathbf{k}) \sin (kx)
    +
    k_{[j} g_{k]} (\mathbf{k}) \cos (kx)
\right] 
\nonumber
\\
& - 
\frac{4 e^{-|\mathbf{k}|}}{c^4 \sqrt{2 \pi}|\mathbf{k}|^2} 
\left[
    \left( 
        k_k k_{(i} f_{j)} (\mathbf{k})
        -
        k_j k_{(i} f_{k)} (\mathbf{k})
    \right) \sin (kx)
    +
    \left( 
    k_k k_{(i} g_{j)} (\mathbf{k}) 
    -
    k_j k_{(i} g_{k)} (\mathbf{k})
    \right) \cos (kx)
\right]
\, .
\label{GEM-knot-wave-Riemann-tensor-1}
\end{align}
The other non-vanishing components of the Riemann tensor can be obtained from the symmetry properties, giving $-R_{i0jk} = -R_{jki0} = R_{0 i j k}$.

The derived components of the Riemann tensor indicate that the curvature induced by the NNTKGMW background is directly related to the wave vector $\mathbf{k}$ and the vectors $\mathbf{f}(\mathbf{k})$ and $\mathbf{g}(\mathbf{k})$, which encode the topological information of the torus knots. This relationship emphasizes the connection between the wave's topological characteristics and the resulting spacetime curvature.
The presence of exponential decay factors $e^{-|\mathbf{k}|}/|\mathbf{k}|^2$ suggests that the influence of the NNTKGMW fields on spacetime curvature diminishes rapidly with increasing wave vector magnitude and frequency This is an expected behavior of gravitational perturbations in the weak-field regime.
The explicit dependence of the Riemann tensor components on trigonometric functions of the wave phase $kx$ shows the oscillatory nature of the induced curvature. This oscillatory behavior is characteristic of monochromatic wave solutions and is a consequence of the periodic oscillations of the underlying torus knot structure.

\subsubsection{The Ricci Tensor}

The Ricci tensor describes the volume distortion of small geodesic balls in the presence of torus knotted monochromatic gravitational waves propagating in the Minkowski background. To calculate its components, we express the Ricci tensor in linearized gravity in terms of the components of the perturbative metric
\begin{equation}
R_{\mu \nu} = {R^{\alpha}}_{\mu \alpha \nu} = 
\frac{1}{2} 
\eta^{\alpha \beta}
\left( 
    h_{\alpha \nu , \mu \beta}
    +
    h_{\alpha \mu , \nu \beta}
\right)
-
\frac{1}{2}
\left( 
    h_{, \mu \nu}
    +
    \partial^2 h_{\mu \nu}
\right)
\, .
\label{GEM-knot-wave-Ricci-tensor}
\end{equation}
After performing standard algebraic calculations, we arrive at the following components of the Ricci tensor
\begin{align}
R_{0 0} 
& = 0
\, ,
\label{GEM-knot-wave-Ricci-1}
\\
R_{0 i} 
& =  
- \frac{e^{-|\mathbf{k}|}}{c^2 \sqrt{2 \pi}}
\left[ 
    f_i (\mathbf{k}) \sin (kx)
    +
    g_i (\mathbf{k}) \cos (kx)
\right] 
\nonumber
\\ 
& +
\frac{4 e^{-|\mathbf{k}|}}{c^4 \sqrt{2 \pi}} 
\left[ 
    f_i (\mathbf{k}) \sin (kx)
    +
    g_i (\mathbf{k}) 
    \cos (kx)
\right]
\, ,
\label{GEM-knot-wave-Ricci-3}
\\
R_{i j} 
& = 
- \frac{2 e^{-|\mathbf{k}|} k_0}{c^2 \sqrt{2 \pi} |\mathbf{k}|^2} 
\left[ 
    k_{(i} f_{j)} (\mathbf{k}) \sin (kx)
    +
    k_{(i} g_{j)} (\mathbf{k}) \cos (kx)
\right] 
\nonumber
\\
& +
\frac{4 e^{-|\mathbf{k}|}}{c^4 k_0 \sqrt{2 \pi}} 
\left[
    k_{(i} f_{j)} (\mathbf{k}) \sin (kx) 
    + 
    k_{(i} g_{j)} (\mathbf{k}) \cos (kx)
\right]
\, .
\label{GEM-knot-wave-Ricci-4}
\end{align}
The above equations show that the NNTKGMW represents an oscillatory perturbation in the spatial and space-time components of the Minkowski geometry, consistent with the nature of such waves in general relativity. The Ricci tensor components $R_{0i}$ and $R_{ij}$ are harmonic functions of $kx$ whose amplitudes decay rapidly with $e^{-|\mathbf{k}|}/|\mathbf{k}|^2$.

Finally, the Ricci scalar of the NNTKGMW geometry can be calculated from the relation
\begin{equation}
R = R^\mu{ }_\mu=\partial^\alpha \partial^\beta h_{\alpha \beta}-
\partial^2 h 
\, .
\label{GEM-knot-wave-Ricci-scalar}
\end{equation}
It is easy to see that the Ricci scalar vanishes $R=0$. This indicates that the NNTKGMW background is flat, which is consistent with the weak field approximation and the absence of sources for the gravitational field.

\subsection{Geodesic Equation}

To describe the dynamics in the NNTKGMW background, we derive the geodesic equation expressed in terms of linearized connection coefficients or Christoffel symbols. From equation (\ref{GEM-knot-wave-metric}), we can obtain the linearized Christoffel symbols, which have the following form in terms of metric perturbations
\begin{equation}
\Gamma_{\mu \nu}^\alpha=\frac{1}{2} \eta^{\alpha \rho}\left(h_{\mu \rho, \nu}+h_{\nu \rho, \mu}-h_{\mu \nu, \rho}\right)
\, .
\label{GEM-knot-wave-Christoffel}
\end{equation}
By reading off the metric components from equation (\ref{GEM-knot-wave-metric}), and after some algebra, we obtain the following Christoffel symbols
\begin{align}
\Gamma^{0}_{00} 
& = 
\Gamma^{0}_{0i}  = 0
\, ,
\label{GEM-knot-wave-Christoffel-1}
\\
\Gamma^{0}_{i j} 
& = 
\left( 1-\frac{2}{c^2} \right)
\frac{e^{-|\mathbf{k}|}}{c^2 \sqrt{2 \pi}|\mathbf{k}|^2}
\left[
    k_{(i} f_{j)}(\mathbf{k}) \cos (kx) 
    -
    k_{(i} q_{j)}(\mathbf{k}) \sin (kx)
\right] 
\, ,
\label{GEM-knot-wave-Christoffel-2}
\\
\Gamma^{i}_{00}
& =
\frac{e^{-|\mathbf{k}|} k_0}{c^2 \sqrt{2 \pi} |\mathbf{k}|^2}
\left[
    f^i (\mathbf{k}) \cos (kx) - g^i(\mathbf{k}) \sin (kx)
\right]
\, ,
\label{GEM-knot-wave-Christoffel-3}
\\
\Gamma^{i}_{0 j}
& = 
\frac{e^{-|\mathbf{k}|}}{c^2 \sqrt{2 \pi}|\mathbf{k}|^2} 
\eta^{ip}\left[
    \left(
        k_{[p} f_{j]}(\mathbf{k})+\frac{2}{c^2} k_{(p} f_{j)}(\mathbf{k})
    \right) \cos (kx)
    -
    \left(
        k_{[p} g_{j]}(\mathbf{k})+\frac{2}{c^2} k_{[p} g_{j]}(\mathbf{k})
    \right) \sin (kx)
\right]
\, ,
\label{GEM-knot-wave-Christoffel-4}
\\
\Gamma^{i}_{j p} 
& = \frac{2 e^{-|\mathbf{k}|}}{c^4 k_0 \sqrt{2}|\mathbf{k}|^2} 
\eta^{iq}
\left\{
    \left[
        k_p k_{(q} f_{j)}(\mathbf{k})+k_j k_{(p} f_{q)}(\mathbf{k})
        -
        k_q k_{(j} f_{p)}(\mathbf{k})
    \right] \cos (kx)
    \right. 
\nonumber
\\
& - 
\left.
    \left[
        k_p k_{(q} g_{j)}(\mathbf{k})
        +
        k_j k_{(p} g_{q)}(\mathbf{k})-k_q k_{(j} g_{p)}(\mathbf{k})
    \right] \sin (kx)
\right\}
\, .
\label{GEM-knot-wave-Christoffel-5}
\end{align} 
From the above relations, we can find the equations of motion of a test particle propagating in the NNTKGMW background. Recall the definition of the geodesic equation
\begin{equation}
\frac{d^2 x^\alpha}{d \lambda^2}+\Gamma_{\mu \nu}^\alpha \frac{d x^\mu}{d \lambda} \frac{d x^\nu}{d \lambda}=0 
\, .
\label{GEM-knot-wave-geodesic}
\end{equation}
After replacing the expressions from the right-hand side of equations (\ref{GEM-knot-wave-Christoffel-1})-(\ref{GEM-knot-wave-Christoffel-5}) into equation (\ref{GEM-knot-wave-geodesic}), the geodesic equation takes the following form in components
\begin{align}
\frac{d^2 x^0}{d \lambda^2} 
& +
\left(1-\frac{2}{c^2}\right)
\frac{e^{-|\mathbf{k}|}}{c^2 \sqrt{2 \pi}|\mathbf{k}|^2}
\left[ k_{(i} f_{j)} (\mathbf{k}) \cos (kx) 
- k_{(i} g_{j)}(\mathbf{k}) \sin (kx) \right] 
\frac{d x^i}{d \lambda} \frac{d x^j}{d \lambda} = 0
\, ,
\label{GEM-knot-wave-geodesic-1}
\\
\frac{d^2 x^i}{d \lambda^2} 
& +
\frac{e^{-|\mathbf{k}|} k_0}{c^2 \sqrt{2 \pi} |\mathbf{k}|^2}
\left[
    f^i(\mathbf{k}) \cos(kx) - g^i (\mathbf{k}) \sin(kx)
\right]
\frac{d x^0}{d \lambda} \frac{d x^0}{d \lambda}
\nonumber
\\
& +
\frac{e^{-|\mathbf{k}|}}{c^2 \sqrt{2 \pi}|\mathbf{k}|^2} 
\eta^{ip}
\left[
    \left(
        k_{[p} f_{j]}(\mathbf{k})+\frac{2}{c^2} k_{(p} f_{j)}(\mathbf{k})
    \right) \cos (kx)
\right.
\nonumber
\\
& -
\left.
    \left(
        k_{[p} g_{j]}(\mathbf{k})+\frac{2}{c^2} k_{[p} g_{j]}(\mathbf{k})
    \right) \sin (kx)
\right]
\frac{d x^0}{d \lambda} \frac{d x^j}{d \lambda}
\nonumber
\\
& + 
\frac{2 e^{-|\mathbf{k}|}}{c^4 k_0 \sqrt{2}|\mathbf{k}|^2} 
\eta^{iq}
\left\{
    \left[
        k_p k_{(q} f_{j)}(\mathbf{k})+k_j k_{(p} f_{q)}(\mathbf{k})
        -
        k_q k_{(j} f_{p)}(\mathbf{k})
    \right] \cos (kx)
    \right. 
\nonumber
\\
& - 
\left.
    \left[
        k_p k_{(q} g_{j)}(\mathbf{k})
        +
        k_j k_{(p} f_{q)}(\mathbf{k})-k_q k_{(j} g_{p)}(\mathbf{k})
    \right] \sin (kx)
\right\}
\frac{d x^j}{d \lambda} \frac{d x^p}{d \lambda} = 0
\, .
\label{GEM-knot-wave-geodesic-2}
\end{align}
Some comments are in order here. Equations (\ref{GEM-knot-wave-geodesic-1}) and (\ref{GEM-knot-wave-geodesic-2}) include nonlinear terms originating from two distinct features: i) the products of derivatives of the coordinates $x^{\mu}$ with respect to the proper time $\lambda$, and ii) the appearance of variables $x^{\mu}$ in the argument of the harmonic functions $\cos(kx)$ and $\sin(kx)$. These non-linearities complicate considerably the general analysis of these equations. However, we observe that a common property of the terms involving the velocity components are the exponential factors $e^{-|\mathbf{k}|}/|\mathbf{k}|^{-2}$. As before, these indicate that the influence of the gravitational wave decreases with the magnitude of the wave vector $|\mathbf{k}|$ or its frequency. The terms involving $k_{(i} f_{j)}$ and $k_{[i} g_{j]}$ represent the force which carries information about the topological properties of the NNTKGMW field. The presence of the harmonic functions $\cos(kx)$ and $\sin(kx)$ indicates that the gravitational wave induces oscillatory forces on the test particle. These forces depend on the wave vector $\mathbf{k}$ and the torus knot represented by $\mathbf{f}(\mathbf{k})$ and $\mathbf{g}(\mathbf{k})$.

The equations (\ref{GEM-knot-wave-geodesic-1}) and (\ref{GEM-knot-wave-geodesic-2}) contain corrections to the equation of a straight line in Minkowski space-time of orders $c^{-5}$, $c^{-4}$, $c^{-3}$, and $c^{-2}$. The dominant effects of the gravitational waves of the NNTKGMW on the dynamics of a test particle are captured by the terms up to $c^{-2}$, for which the equations of motion take the following form
\begin{align}
\frac{d^2 x^0}{d \lambda^2} 
& +
\frac{e^{-|\mathbf{k}|}}{c^2 \sqrt{2 \pi}|\mathbf{k}|^2}
\left[ k_{(i} f_{2j)} (\mathbf{k}) \cos (kx) 
- k_{(i} g_{2j)}(\mathbf{k}) \sin (kx) \right] 
\frac{d x^i}{d \lambda} \frac{d x^j}{d \lambda} = 0
\, ,
\label{GEM-knot-wave-geodesic-1-simplified-1}
\\
\frac{d^2 x^i}{d \lambda^2} 
& +
\frac{e^{-|\mathbf{k}|} k_0}{c^2 \sqrt{2 \pi} |\mathbf{k}|^2}
\left[
    f^i_2(\mathbf{k}) \cos (kx)
    -
    g^i_2 (\mathbf{k}) \sin (kx)
\right]
\frac{d x^0}{d \lambda} \frac{d x^0}{d \lambda}
\nonumber
\\
& +
\frac{e^{-|\mathbf{k}|}}{c^2 \sqrt{2 \pi}|\mathbf{k}|^2} 
\eta^{ip}
\left[
    k_{[p} f_{2j]}(\mathbf{k}) \cos (kx)
    -
    k_{[p} g_{2j]}(\mathbf{k}) \sin (kx)
\right]
\frac{d x^0}{d \lambda} \frac{d x^j}{d \lambda}
= 0
\, , 
\label{GEM-knot-wave-geodesic-2-simplified-2}
\end{align}
where
\begin{equation}
\mathbf{f}_2 (\mathbf{k}) = \begin{bmatrix}
    - n \vert \mathbf{k} \vert k^2
    \\
    n \vert \mathbf{k} \vert k^1
    \\
    0
\end{bmatrix}
\, ,
\qquad
\mathbf{g}_2 (\mathbf{k}) = \begin{bmatrix}
    - m k^1 k^3
    \\
    - m k^2 k^3
    \\
    m \left[ (k^1)^2+(k^2)^2 \right]
\end{bmatrix}
\, .
\label{GEM-knot-wave-geodesics-f2g2}
\end{equation}

The equations (\ref{GEM-knot-wave-geodesic-1-simplified-1}) and (\ref{GEM-knot-wave-geodesic-2-simplified-2}) are still highly non-linear. However, they show that, at leading order, a particle's dynamics is determined by the gravitomagnetic component of the NNTKGMW, as can be seen from the vectors $\mathbf{f}_2 (\mathbf{k})$ and $\mathbf{g}_2 (\mathbf{k})$, which contain information about the gravitomagnetic field only. As before, the gravitational wave effect is manifest in the terms containing the harmonic functions $\cos(kx)$ and $\sin(kx)$. Their presence in the equations displays the oscillatory nature of the gravitational waves and their effect on the test particle's trajectory. Finding non-trivial solutions to these equations is beyond the scope of the present work.

\section{GEM Properties of NNTKGMW}

The NNTKGMW fields discussed before have specific physical and mathematical properties that are the consequence of their origin in the GEM formalism. In this section, we discuss two of these properties, namely the dual NNTKGMW geometry and the GEM helicity.

\subsection{Dual NNTKGMW Geometry}

The GEM equations obey the electric-magnetic duality of Maxwell's equation in vacuum. Since the GEM potentials are interpreted as GEM components of the perturbative metric, the potentials of the dual GEM fields correspond to the GEM components $\tilde{\mathsf{h}}_{0\mu}$ of the dual perturbative metric $\tilde{\mathsf{h}}_{\mu\nu}$. Here, tilde denotes the trace-reversed transformation of $\mathsf{h}_{\mu\nu}$ as before. Although there is still no consensus on the interpretation of this dual geometry, it is interesting to calculate the dual GEM background of the NNTKGMW. In particular, the dual potentials are useful to calculate GEM physical objects such as the gravitomagnetic and gravitoelectric helicities, by analogy with classical electrodynamics\footnote{A possible insight into the interpretation of the dual GEM metric could be gained from interesting connections between the electric-magnetic duality of Ra\~{n}ada fields and the Finsler geometry of classical charges, as discussed in \cite{Crisan:2020krc}.}.

The electric-magnetic duality in GEM can be established as in classical electrodynamics by noting that the GEM equations (\ref{GEM-knot-wave-GEM-1}) and (\ref{GEM-knot-wave-GEM-2}) are invariant under the mapping
\begin{equation}
\mathbf{E} \mapsto c \mathbf{B}
\, ,
\qquad
c \mathbf{B} \mapsto -\mathbf{E}
\, .
\label{GEM-knot-wave-duality-1}
\end{equation}
Therefore, there are GEM potentials $(\Psi, \mathbf{C})$, such that
\begin{equation}
\mathbf{E} = \nabla \times \mathbf{C}, 
\qquad 
\mathbf{B}=\nabla \Psi + c^{-2} \partial_t \mathbf{C} 
\label{GEM-knot-wave-duality-2}
\, .
\end{equation}
Since $\tilde{\mathsf{h}}_{\mu\nu}$ are required to be solutions of the NNTKGMW type of the linearized Einstein's equations in vacuum, one must impose the same gauge conditions as on $\tilde{h}_{\mu\nu}$. In particular, we set $\Psi = 0$ and $\nabla \cdot \mathbf{C} = 0$. In this gauge, we can express the GEM fields in terms of both potentials $\mathbf{A}$ and $\mathbf{C}$ as follows
\begin{align}
\mathbf{B} & = \nabla \times \mathbf{A} = \frac{1}{c^2} \frac{\partial \mathbf{C}}{\partial t}, 
\label{GEM-knot-wave-duality-3}
\\
\mathbf{E} & = \nabla \times \mathbf{C} = -\frac{\partial \mathbf{A}}{\partial t} 
\label{GEM-knot-wave-duality-4}
\, .
\end{align}
It is easy to see that the dual NNTKGMW has the form
\begin{equation}
C^j (x;k )= 
c^j (\mathbf{k}) e^{-i k x} 
\, .
\label{GEM-knot-wave-dual-wave}
\end{equation}
where the components of the amplitude vector $\mathbf{c} (\mathbf{k})$ are related to the Fourier components of the non-null torus knotted GEM fields by the relation
\begin{equation}
c^j (\mathbf{k}) = \frac{-i}{2 c | \mathbf{k}| } 
\left(
    c \mathcal{B}^j (\mathbf{k} )
    +
    \varepsilon^{j}_{pq} k^p \mathcal{E}^q (\mathbf{k})
\right).
\end{equation}
Using equations (\ref{GEM-knot-wave-E-Fourier-non-null-knot}) and (\ref{GEM-knot-wave-B-Fourier-non-null-knot}), the same algebra as before gives
\begin{equation}
\mathbf{C}(x;k) = -\frac{i e^{-|\mathbf{k}|} e^{-ikx}}{2 \sqrt{2 \pi}|\mathbf{k}|^2 }
\left[
\begin{array}{c}
c n k^1 k^3 + i \left[c m |\mathbf{k}| k^2 + s \left[(k^2)^2 + (k^3)^2 \right] \right] \\
c n k^2 k^3 + l |\mathbf{k}| k^3 - i \left[c m |\mathbf{k}| k^1 + s k^1 k^2 \right] \\
-c n \left[(k^1)^2 + (k^2)^2 \right] - l |\mathbf{k}| k^2 - i s k^1 k^3
\end{array}
\right]
\, .
\label{GEM-knot-wave-dual-c}
\end{equation}
Equation (\ref{GEM-knot-wave-dual-c}) represents the components of the GEM sector $\tilde{\mathsf{h}}_{0 i}$ of the dual metric perturbation. It is easy to verify that the orthogonality condition $k_i C^i (x;k) = 0$ is satisfied. In the spatial sector, we can construct $\tilde{\mathsf{h}}_{ij}$ as in equation (\ref{GEM-knot-wave-sol-gamma}), namely
\begin{equation}
\chi_{ij} (x;k )= 
\upchi_{ij} (\mathbf{k}) e^{-i k x} 
\, .
\label{GEM-knot-wave-dual-c-gamma}
\end{equation}
In order to construct the spatial components of the dual metric, we make the same ansatz from equation (\ref{GEM-knot-wave-sol-gamma}) and impose the symmetry and traceless conditions $\chi_{ij} = \chi_{ji}$ and $\text{Tr} (\chi_{ij}) = 0$, and we obtain
\begin{equation}
\chi_{ij} (x;k )= 
k_{(i} c_{j)}(\mathbf{k})  \, \frac{e^{-i k x}}{k_0}  
\, .
\label{GEM-knot-wave-sol-dual-c-gamma-k}
\end{equation}
The construction of the dual NNTKGMW perturbative metric follows exactly the same steps as before, with $\tilde{\mathsf{h}}_{\mu\nu} = \{ \tilde{\mathsf{h}}_{00}, \tilde{\mathsf{h}}_{0i} = \tilde{\mathsf{h}}_{i0}, \tilde{\mathsf{h}}_{ij} \} \sim \{ 0, 2C_i, 2\chi_{ij}\}$. To obtain a real line element, we separate the real part of it which is given by
\begin{align}
C^{i}_R(x; \mathbf{k})
& = 
-\frac{e^{-|\mathbf{k}|}}{2 \sqrt{2 \pi} |\mathbf{k}|^2}
\left[ 
\tilde{f}^{i}(\mathbf{k}) \sin (kx) - \tilde{g}^i (\mathbf{k}) \cos (kx)
\right]
\, ,
\label{GEM-knot-wave-real-dual-1}
\\
[\chi_R]_{ij} (x;k )
& =
- \frac{e^{-|\mathbf{k}|}}{2 k_0 \sqrt{2 \pi} |\mathbf{k}|^2}
\left[ 
k_{(i} \tilde{f}_{j)}(\mathbf{k}) \sin (kx)
-
k_{(i} \tilde{g}_{j)}(\mathbf{k}) \cos (kx)
\right]
\, ,
\label{GEM-knot-wave-real-dual-2}
\end{align}
where we have introduced the following dual vector amplitudes:
\begin{align}
\tilde{\mathbf{f}} (\mathbf{k})
&=
\frac{1}{c}
\left[
\begin{array}{l}
c n k^1 k^3 \\
c n k^2 k^3 + l|\mathbf{k}| k^3 \\
- c n \left[(k^1)^2 + (k^2)^2 \right] - l|\mathbf{k}| k^2
\end{array}
\right]
\, ,
\label{GEM-knot-wave-real-dual-3}
\\
\tilde{\mathbf{g}} (\mathbf{k})
&=
\left[
\begin{array}{l}
c m|\mathbf{k}| k^2 + s\left[(k^2)^2 + (k^3)^2 \right] \\
- c m|\mathbf{k}| k^1 - s k^1 k^2 \\
- s k^1 k^3
\end{array}
\right]
\, .
\label{GEM-knot-wave-real-dual-4}
\end{align}
Then the metric of the dual NNTKGMW geometry can be written as
\begin{align}
d \tilde{s}^2 (k)  
& = 
(dx^0)^2
-  
\frac{e^{- \vert \mathbf{k} \vert}}{c^2 \sqrt{2 \pi} \vert \mathbf{k} \vert^2}
\left[
\tilde{f}_i (\mathbf{k}) \sin (kx) - \tilde{g}_i (\mathbf{k}) \cos (kx)
\right]
dx^0 dx^i
\nonumber
\\
& -
\left[\eta_{i j} + 
\frac{e^{- \vert \mathbf{k} \vert}}{c^4 \sqrt{2 \pi} k_0 \vert \mathbf{k} \vert^2}
\left[
k_{(i} \tilde{f}_{j)} (\mathbf{k}) \sin (kx)
-
k_{(i} \tilde{g}_{j)} (\mathbf{k}) \cos (kx)
\right]
\right] d x^i d x^j
\, .
\label{GEM-knot-wave-metric-dual}
\end{align}

As we can see from (\ref{GEM-knot-wave-metric-dual}), the dual NNTKGMW geometry has a very similar structure to the NNTKGMW geometry, from which we conclude that it has similar properties. If an overall negative sign is absorbed into the definition of $\tilde{\mathbf{f}} (\mathbf{k})$, then the geometric objects of the dual geometry, that is, the Christoffel symbols, Riemann tensor, Ricci tensor, Ricci curvature, and the geodesic equation, can be immediately written down by simply replacing non-tilde with tilde vectors in the corresponding equations. Since this is a very simple exercise, we leave it to the reader.

\subsection{GEM Helicities of NNTKGMW Background}

The main utility of calculating the dual potentials is that they allow one to construct the GEM helicity. In electrodynamics, the helicity of both electric and magnetic components is important as they form a constant of motion for the electromagnetic field in vacuum \cite{Trueba:1996}, and thus are relevant for the study of the dynamics of the torus knotted electromagnetic fields \cite{Arrayas:2011ci,Arrayas:2016bqn}. However, the GEM helicity is a more controversial subject. As discussed in \cite{Bini:2021gdb}, the gravitomagnetic helicity can be used as a measure of the complexity of the gravitomagnetic fields present in a large variety of gravitational and astrophysical processes. On the other hand, gravitoelectric fields do not seem to be directly connected to any phenomenology to date. Nevertheless, in this subsection, we provide both gravitomagnetic and gravitoelectric helicities since the calculations in both cases are identical.

In analogy with electrodynamics, the GEM helicities are defined as
\begin{align}
H_m (k) & =  - \int d^3 x \, A_{Ri} B^{i}_{R} =
- \int d^3 x \varepsilon_{ijp} A^{i}_{R} \partial^j A^{p}_{R} 
\, ,
\label{GEM-knot-wave-hel-m}
\\
H_e (k)& = - \int d^3 x \, C_{Ri} E^{i}_{R} =
- \int d^3 x \varepsilon_{ijp} C^{i}_{R} \partial^j C^{p}_{R} 
\, ,
\label{GEM-knot-wave-hel-e}
\end{align}
where the index $R$ stands for the real quantities calculated before and the integral is taken over the spatial volume. The calculations use the gravitomagnetic and gravitoelectric fields as a function of GEM potential. Since this relation is a consequence of the GEM equations, in this sense the above relations are defined on-shell. By substituting equations (\ref{GEM-knot-wave-sol-A-real}) and (\ref{GEM-knot-wave-real-dual-1}) into equations (\ref{GEM-knot-wave-hel-m}) and (\ref{GEM-knot-wave-hel-e}), respectively, and after some straightforward algebra, we obtain
\begin{align}
H_m (k) & = \frac{e^{-2|\mathbf{k}|}}{8 \pi |\mathbf{k}|^4} \varepsilon_{i j p} f^i (\mathbf{k}) k^j g^p (\mathbf{k}) V_3
\, ,
\label{GEM-knot-wave-hel-m-1}
\\
H_e (k) & = \frac{e^{-2|\mathbf{k}|}}{8 \pi |\mathbf{k}|^4} \varepsilon_{i j p} \tilde{f}^i (\mathbf{k}) k^j \tilde{g}^p (\mathbf{k}) V_3
\, ,
\label{GEM-knot-wave-hel-e-2}
\end{align}
where $V_3$ is the three-dimensional volume. Since the helicity densities are finite, the helicities diverge for large volumes. 
As equations (\ref{GEM-knot-wave-hel-m}) and (\ref{GEM-knot-wave-hel-e}) show, the helicities measure the interaction between the GEM potentials and the corresponding GEM fields in the GEM sector of the perturbative metric. However, since the same potentials are used to construct the spatial components too, we naively expect that the latter also interact with the GEM fields. We propose as the simplest and most natural measure of this interaction the following generalization of GEM helicity to the spatial sector
\begin{align}
H^{(s)}_m = \int d^3 x \left[ \mathrm{\gamma}_R \right]_{ij} (x;k ) B^{i}_{R} B^{j}_{R}
\, ,
\label{GEM-knot-wave-hel-space-m}
\\
H^{(s)}_e = \int d^3 x \left[ \mathrm{\chi}_R \right]_{ij} (x;k ) C^{i}_{R} C^{j}_{R}
\, .
\label{GEM-knot-wave-hel-space-e}
\end{align} 
Again, by substituting the GEM fields in terms of their potentials and performing the algebra, we obtain
\begin{equation}
H^{(s)}_m = H^{(s)}_e = 0
\, .
\label{GEM-knot-wave-hel-space-0}
\end{equation} 
The above result shows that the spatial component of the NNTKGMW does not have GEM helicity in the sense defined above. It is possible that, in order to study this property, more refined observables may be necessary for the NNTKGMW background.

\section{Discussions}

In this paper, we have constructed a new non-null torus knotted gravitational monochromatic wave (NNTKGMW) solution of Einstein linearized equations in vacuum within the extended GEM formalism. To obtain this solution, we have fixed the gauge symmetry of the linearized equations and the gauge symmetry of GEM equations as in classical electrodynamics. 

The NNTKGMW solution has several interesting properties. Firstly, its amplitude carries information about the torus on which the gravitoelectric and gravitomagnetic fields wind up at the initial time. Secondly, the solution is constructed for all components of the perturbed metric from  both the GEM sector and the spatial sector. Thirdly, the NNTKGMW solution describes a family of gravitational waves parameterized by four integers that describe the winding numbers of the gravitoelectric and gravitomagnetic fields around the torus cycles and the three real numbers that are the components of the wave vector. Two real parameters, which are the torus radii and whose magnitudes are gauge-dependent, can be added to the NNTKGMW fields.

Next, we have addressed the problem of the geometry of the NNTKGMW background by calculating the relevant geometrical objects: the Riemann tensor, the Ricci tensor, and the Ricci scalar. Also, we derived the geodesic equation. On components, it is represented by a system of highly non-linear differential equations whose non-linearity has two sources: the standard product of velocity components and the presence of the coordinate components in the arguments of harmonic functions. At the leading order $c^{-2}$, the geodesic equation is dominated by the gravitomagnetic component, but the non-linearities persist.

The NNTKGMW has specific properties that result from its GEM origin. We have discussed two such properties: the dual NNTKGMW solution and its geometry, which results from the electric-magnetic duality of GEM equations, and the GEM helicity. Based on the interpretation of the latter as the integrated interaction between the GEM potential and its corresponding field and on the fact that the NNTKGMW amplitude in the spatial sector of the perturbative metric is linear in the GEM potentials, we have proposed a simple generalization of the GEM helicity to the spatial sector. This quantity vanishes, which shows that the spatial sector of metric has no GEM helicity as defined.

The NNTKGMW presented here is interesting for several reasons. From a conceptual viewpoint, it represents a GEM solution for all components of the perturbative metric and adds to the body of solutions in the standard GEM approach to the linearized Einstein equations. From a physical viewpoint, the NNTKGMW provides a new type of gravitational wave carrying topological information in its amplitudes. However, the generation of these waves is not clear, as they have been restricted to the vacuum equations and to a specific gauge equivalent to the TT-gauge. Nevertheless, they demonstrate that the linearized gravitational vacuum can carry information about topologically non-trivial objects. This result is consistent with similar conclusions drawn in works on Weyl's approach to GEM, as mentioned in the introduction.

Therefore, this work opens up several interesting lines of research. The first consists of further investigating the physical properties of NNTKGMW solutions and attempting to connect them to phenomenology in relevant applications. A second line of research is to study the non-linear geodesic equations in the search for non-trivial solutions. A third line involves the search for an interpretation of the dual geometry. 

Another important problem is related to the fact that
the non-null torus knotted gravitoelectric and gravitomagnetic fields discussed exhibit non-trivial topological structures encoded in the wave amplitudes and that can be characterized by their winding numbers $(m,n)$ and $(l,s)$. The presence of these topological invariants suggests potential implications for understanding the interplay between topology and gravitational waves. Future work may explore the stability and evolution of these topological structures under various perturbations.

In applying the extended GEM formalism in this work, very specific gauge conditions have been imposed to obtain knotted wave solutions: the choice of the transverse gauge and the imposition of $\Phi = 0$ were crucial in deriving the presented solutions. Investigating other gauge choices and their implications on the solutions is an interesting direction for future research.

The current work focuses on monochromatic wave solutions with specific topological configurations. Future studies could generalize these results to include larger classes of solutions, such as linear superposition 
of NNTKGMW in both linearized classical and quantum gravity or other knotted structures. Investigating the interactions between multiple knotted gravitational waves and their collective dynamics could provide further insights into the nonlinear behavior of gravitoelectromagnetic fields and of the perturbative metric.

\section*{Acknowledgments}
We would like to thank to A. V. Crisan, C. F. L. Godinho, M. C. Rodriguez and D. T. Concei\c{c}\~{a}o Jr. for fruitful discussions. 
I. V. Vancea received support from the Basic Research Grant (APQ1) from the Carlos Chagas Filho Foundation for Research Support of the State of Rio de Janeiro (FAPERJ), grant number E-26/210.511/2024.


\end{document}